# Competition between excitonic insulators and quantum Hall states in correlated electron-hole bilayers

Ruishi Qi[1,†], Qize Li[1,†], Zuocheng Zhang[1], Zhiyuan Cui[1], Bo Zou[2], Haleem Kim[1,3], Collin Sanborn[1,3,4], Sudi Chen[1], Jingxu Xie[1,3,4], Takashi Taniguchi[5], Kenji Watanabe[6], Michael F. Crommie[1,3], Allan H. MacDonald[2], Feng Wang[1,3,7,*]

[1] Department of Physics, University of California, Berkeley, CA 94720, USA.

[2] Department of Physics, University of Texas at Austin, Austin, TX 78712, USA.

[3] Materials Sciences Division, Lawrence Berkeley National Laboratory, Berkeley, CA 94720, USA.

[4] Graduate Group in Applied Science and Technology, University of California, Berkeley, CA, USA.

[5] Research Center for Materials Nanoarchitectonics, National Institute for Materials Science, 1-1 Namiki, Tsukuba 305-0044, Japan.

[6] Research Center for Electronic and Optical Materials, National Institute for Materials Science, 1-1 Namiki, Tsukuba 305-0044, Japan.

[7] Kavli Energy NanoSciences Institute, University of California Berkeley and Lawrence Berkeley National Laboratory, Berkeley, CA 94720, USA.

[†] These authors contributed equally to this work.

* To whom correspondence should be addressed: fengwang76@berkeley.edu.

**Excitonic insulators represent a unique quantum phase of matter, providing a rich ground for studying exotic quantum bosonic states. Strongly coupled electron-hole bilayers, which host stable dipolar exciton fluids with an exciton density that can be adjusted electrostatically[1–4], offer an ideal platform to investigate correlated excitonic insulators[5,6]. Based on electron-hole bilayers made of $MoSe_2$/hBN/$WSe_2$ heterostructures, here we study the behavior of excitonic insulators in a perpendicular magnetic field. We report the observation of excitonic quantum oscillations in both Coulomb drag signals and electrical resistance at low to medium magnetic fields. Under a strong magnetic field, we identify multiple quantum phase transitions between the excitonic insulator phase and the bilayer quantum Hall insulator phase. These findings underscore the interplay between the electron-hole interactions and Landau level quantization that opens new possibilities for exploring quantum phenomena in composite bosonic insulators.**

Electron-hole (e-h) bilayers, where a two-dimensional electron gas (2DEG) and a 2D hole gas (2DHG) are confined to two closely spaced but electrically isolated layers[7,8], provide a versatile platform to study quantum phases of correlated e-h fluids. This configuration facilitates strong Coulomb interactions between the electrons and holes, leading to the formation of bound exciton states. The spatial separation between the layers suppresses e-h recombination, enabling long-lived exciton fluids in thermal equilibrium that can be controlled electrostatically. These systems are of great interest for exploring novel quantum bosonic and fermionic phases, including excitonic insulators (EIs)[1,2,7], exciton Bose-Einstein condensates[8–13], and higher-order e-h complexes such as trions and biexcitons[14–17].

Recent advances have allowed for the creation of strongly coupled e-h bilayers using transition metal dichalcogenide (TMD) heterostructures[1–4,16,17]. With large effective mass and reduced dielectric screening in TMD systems, the strong Coulomb attraction between electrons and holes across the layers leads to spontaneous formation of interlayer excitons. Compared with previous system based on III-V semiconductor quantum wells or graphene systems[11,18–25], the strong e-h coupling regime is accessible without requiring external magnetic fields. A clearly-defined EI ground state at zero magnetic field has been reported, featuring robust exciton binding, a large charge gap (~20 meV)[1,2] and perfect Coulomb drag[3,4]. Theoretical studies suggested that when an out-of-plane magnetic field is applied to these excitonic insulating e-h bilayers, the interplay between Coulomb interactions and magnetic cyclotron energy can produce quantum oscillations (QOs) within the EI phase[5,6]. In the strong field limit, it is also predicted that the magnetic field can destroy the exciton binding and turn the EI into two independent quantum Hall insulators (QHIs)[5]. The presence of QOs – a defining feature of a metallic system – in correlated insulating phases is intriguing, and has been a topic under debate in the past decade[26–39]. Experimental observations of the predicted excitonic QOs and quantum phase transitions have remained elusive.

In this paper, we report on the first experimental study of excitonic insulating e-h bilayers under an out-of-plane magnetic field in MoSe$_2$/hBN/WSe$_2$ heterostructures. At zero magnetic field, the EI phase is strongly insulating and exhibits perfect Coulomb drag behavior, in which a drive current in one layer induces an equal amount of drag current in the second layer. As a magnetic field is applied, we observe clear QOs of the drag current and the electrical resistance. Our findings demonstrate unambiguously that QOs can emerge in an EI, in striking contrast to the absence of QOs in conventional insulators. At higher magnetic fields, we further observe multiple quantum phase transitions between EI phases, in which layer-separated electrons and holes remain bound as excitons, and bilayer QHI phases, in which two layers form independent quantum Hall states.

## Excitonic insulating electron-hole bilayers

Here we fabricate strongly coupled e-h bilayer devices based on MoSe$_2$/hBN/WSe$_2$ heterostructures, as illustrated in Fig. 1a. The device is composed of two TMD monolayers that act as an electron layer (MoSe$_2$) and a hole layer (WSe$_2$), which are separated by an ultrathin hBN tunneling barrier. They are further sandwiched by graphite top gate (TG) and bottom gate (BG) with hBN dielectric layers. With dual gates and two individually contacted TMD layers, the doping densities of electrons and holes can be independently controlled electrostatically. The MoSe$_2$ conduction band and WSe$_2$ valence band form a type-II band gap of approximately 1.5 eV. The charge gap can be electrically closed by a combination of the interlayer bias voltage $V_B \equiv V_h - V_e$ and the vertical electric field $V_{BG} - V_{TG}$. We apply a large and fixed vertical electric field, and use $V_B$ to continuously tune the charge gap. The symmetric gate voltage $V_G \equiv V_{TG}/2 + V_{BG}/2$ controls the Fermi level and therefore the e-h density imbalance.

To perform transport measurements, we fabricate two charge reservoir regions with increased interlayer distance on the two sides of the active region of interest. In the reservoir regions, the vertical electric field induces a much larger interlayer voltage difference due to increased interlayer spacing, creating a heavily doped electron-hole plasma (EHP) that serves as good contacts for excitons[1,2]. Each reservoir region is equipped with one graphite contact to the MoSe$_2$ layer, and three platinum (Pt) electrodes for the WSe$_2$ layer. Fig. 1b shows the optical image of a representative device D1 that has a 2.3 nm thick (7-layer) hBN tunneling barrier. We will focus on this device in the main text, but similar results can be reproduced in another device D2 (Extended Data Fig. 1).

We first characterize the device at zero magnetic field. Taking advantage of the sensitive doping dependence of optical absorption spectrum in monolayer TMDs, the charge doping phase diagram of the e-h bilayer can be determined using optical spectroscopy (see ref.[1]). Fig. 1c-d shows the experimentally determined hole density $n_h$ and electron density $n_e$ as a function of $V_G$ and $V_B$. The constant vertical electric field $\approx 0.45$ V/nm applied here reduces the type-II bandgap from 1.5 eV to approximately 0.37 eV. The interlayer bias voltage $V_B$ controls the exciton chemical potential, and hence the effective charge gap $E_g = 0.37$ eV $- V_B$. At $V_B < 0.34$ V, the system is a normal insulator (NI) with a finite band gap in which the symmetric gate voltage can tune the system from intrinsic to either electron doped or hole doped. With a larger $V_B$ ($0.34 - 0.37$ V), the effective charge gap $E_g$ becomes smaller than the interlayer exciton binding energy but still remains non-zero. The e-h bilayer is known to host an EI state at net charge neutrality, which is exciton-compressible but charge-incompressible[1–4]. Increasing $V_B$ even further destroys the EI through an interaction-driven exciton Mott transition into an EHP phase[1–3].

We employ two types of transport measurements to study the exciton behavior. Unless otherwise specified, the transport measurements below are performed at an electronic temperature $T = 0.5$ K. The first measurement is a two-terminal resistance measurement for the WSe$_2$ layer, with the electron layer open-circuit (Fig. 1e). We utilize two separate pairs of Pt electrodes for sourcing current and measuring voltage, eliminating the Pt-WSe$_2$ junction resistance in the measurement (Methods). Each highly doped reservoir region acts as one terminal for the region of interest. The measured hole-layer resistance $R_h$ characterizes the charge transport behavior of doped holes. A 2D color plot of $R_h$ is provided in Fig. 1f. As expected, the $R_h$ plot shares a similar overall shape with the $n_h$ plot in Fig. 1c, being very resistive when $n_h = 0$. $R_h$ quickly drops to a few kΩ when holes are doped into the system, with the exception of the EI phase, in which the holes remain very resistive at finite $n_h$. This is because in the EI all the holes spontaneously pair with the electrons into the exciton form, leaving no unpaired mobile charge.

The second measurement is the closed-circuit Coulomb drag experiment illustrated in Fig. 1g. A drive current $I_{\text{drive}}$ is passed through the electron layer, while the closed circuit drag current in the WSe$_2$ layer $I_{\text{drag}}$ is measured. When all electrons and holes pair into excitons, the motion of an electron must be accompanied by a hole, leading to perfect Coulomb drag behavior. The drag ratio in Fig. 1h, defined as $I_{\text{drag}}/I_{\text{drive}}$, reaches unity in the EI phase (see Extended Data Fig. 2 for the drive and drag currents). The large $R_h$ value and perfect drag behavior are both unambiguous evidence of the EI phase.

## Electron-hole fluids in a strong magnetic field

We next examine the behavior of the e-h bilayer in a strong perpendicular magnetic field. Fig. 2a-b shows the hole resistance and the drag ratio at $B = 12$ T. Compared with the zero-field case, the most noticeable difference here is the emergent quasi-periodic structures as the density is varied.

In the general hole-doped region, $R_h$ exhibits strong QOs. Fig. 2c displays a vertical linecut of $R_h$ at a fixed $V_G = 0.1$ V (blue dashed line in Fig. 2a), where accurately quantized plateaus at $h/Ne^2$ ($N$ denotes an integer; $h$ and $e$ are the Planck constant and the elementary charge respectively) can be observed. Between adjacent plateaus, the resistance first has a slight increase, followed by a sharper decrease and transition to the next plateau. Meanwhile, there is negligible drag signal. This behavior is consistent with a typical 2DHG in the integer QH regime[40], where the two-terminal resistance is quantized to the Hall resistance when the contact resistance between the 2DHG and the highly doped WSe$_2$ terminals is sufficiently small. We assign the quantized plateaus as integer filling of hole Landau levels (LLs) $\nu_h = 1, 2, \ldots, 6$, as labeled in Fig. 2a-b. Apparently, the 2DHG transport is only weakly influenced by the electrons in the nearby layer.

Although the interlayer coupling effects seem to be weak for the general $n_e \neq n_h$ case, the behavior near net charge neutrality $n_e = n_h$ displays strikingly different features. The triangular EI region in Fig. 1f and 1h now develops into a series of lobes centered along the charge neutral line (CNL, black dotted line in Fig. 2b). These lobes are characterized by an increased $R_h$ value and strong drag signal. The neighboring lobes are connected at low $V_B$, and become separated at higher pair densities. Within each lobe, the drag effect is strongest along the central CNL; with finite e-h imbalance, it gradually decreases outward, and eventually drops to nearly zero. Fig. 2d is a linecut of $R_h$ and drag ratio approximately tracing constant hole filling factor $\nu_h = 3$ (magenta dashed line in Fig. 2a). A large hole resistance $R_h$ and a large drag ratio with a peak value of 0.8 are observed in the charge neutral region. Away from charge neutrality, the drag signal decreases quickly and becomes negligible, while $R_h$ recovers its quantized value at $h/3e^2$, indicating a robust hole QH state when electron density does not match the hole density. Similar behavior is observed at other integer filling factors, as seen in Fig. 2a-b.

Now we focus on the behaviors along the CNL. Fig. 2e displays the CNL linecut of the drag ratio and $R_h$. With increasing $V_B$ (and correspondingly higher exciton density), the drag ratio starts from a plateau at unity, followed by multiple dips at integer LL filling with increasing depth, until the pair density is too high for excitons to bind ($V_B > 0.46$V). For small integer filling factors ($\nu = 1 - 4$), the drag signal does not completely vanish at the dips, and $R_h$ deviates from the ideal quantized value of $h/\nu e^2$. Both indicate LL mixing caused by the strong e-h attraction. For the $\nu = 5$ dip, the drag ratio drops to zero, suggesting complete destruction of excitons where electrons and holes form independent QHIs. Here the $R_h$ value becomes quantized at $h/5e^2$, consistent with QH states with dissipationless edge conducting channels and insulating bulk regions.

Between adjacent QHI phases, strong exciton drag is observed at partial filling of LLs, with peaks at approximately half-integer LL fillings. The drag current from the excitons in partially filled LLs is smaller than but still comparable to the drive current. A qualitative explanation can be built considering the quantized edge conducting channels competing with bulk exciton transport. The drive current $I_{\text{drive}}$ is shared by two parallel paths, the exciton motion and the edge channels of the electron layer. The former induces an equal but opposite current in the hole layer. This induced current in the hole layer does not necessarily all go through the external measurement circuit due to finite contact resistance (estimated to be 2.4 k$\Omega$, Extended Data Fig. 3). A portion of it circulates back to the other end through the edge channels in the hole layer, while the remaining is measured in $I_{\text{drag}}$. Because the edge channel resistance for both layers decreases approximately as $h/\nu e^2$ for increasing LLs, a larger fraction of current passes through the edge channels, and the observed drag ratio decreases.

The QOs and phase transitions can be understood as oscillations of the effective charge gap which is lowered by Landau quantization when the exciton filling factor is close to an integer[5,6]. The dependence of charge gap, the energy difference between the highest occupied dressed LL and the lowest occupied dressed LL, on $V_B$ therefore develops a QO variation on top of the smooth drop present at $B = 0$. The oscillation of charge gap need not fully destroy the EI, only leading to the oscillation of drag ratio and $R_h$ within the EI phase. For larger pair densities, exciton binding weakens due to the many-body effects responsible for the Mott transition. In this regime, it is therefore possible to dissociate all the excitons at some positions within the gap energy oscillation, resulting in an alternation between bilayer QHI phases and strongly correlated EI phases. Note that in our device geometry the carrier densities in the graphite gates (on the order of $10^{13}$ cm$^{-2}$) are always much higher than $n_e$ and $n_h$ (Extended Data Fig. 4). We therefore rule out the possibility of graphite gate induced potential oscillations[41] (which otherwise would happen at a very different frequency), and conclude that the QOs in both drag signals and electrical resistance come from the competition between Landau level quantization and exciton binding.

Our observation of EI and QHI transitions along the CNL agrees qualitatively with the theoretical predictions in refs.[5,6]. The experimentally observed critical magnetic field, however, is much lower, presumably due to an overestimation of exciton binding energy in the theory. In addition, the magnetic EI is present only along the CNL. The e-h Coulomb attraction in TMD heterostructure is very strong and can mix different LLs effectively. Theory predicts that the EI states can also exist when the electron and hole filling factors $\nu_e, \nu_h$ are different by one (or by other integers)[6], but they are not observed experimentally. Further theoretical studies are required to fully describe the competition of EI and QHIs in strongly interacting e-h fluids with a net non-zero charge density.

**Fan diagram at net charge neutrality**

The relative values of the exciton binding energy and the LL quantization energy can be tuned by varying exciton density or $B$ field. As a result, QOs and phase transitions can take place as a function of both $V_B$ and $B$. Fig. 3a shows the $B$ field dependence of the drag ratio along the CNL. The NI, EI and EHP phases have distinct drag ratio characteristics under the magnetic field. In the NI phase at low $V_B$, neither a drive current nor a drag current can be established in the system. At intermediate $V_B$, the EI phase leads to large drag ratios. With increasing magnetic field, the EI region survives to a larger $V_B$, or a larger pair density, due to a reduction of screening effects and increased exciton binding energy[42]. Within the EI region, QOs of the drag ratio start to show up when $B \gtrsim 5$ T, forming Landau-fan-like structures. In the high-density EHP phase, a small frictional drag signal oscillates weakly with magnetic field. Corresponding behaviors also appear in the $R_h$ measurement data shown in Fig. 3b: the NI phase remains resistive regardless of $B$, the resistance in the EI phase forms Landau-fan-like structures that also appear in its drag signal, and the EHP phase develops QH resistance plateaus at high field.

Fig. 3c displays representative linecuts of $R_h$ and the drag ratio at three $V_B$ values, plotted against $1/B$. At $V_B = 0.35$ V, the exciton binding energy at low pair density (>20 meV) is much larger than the magnetic energy scale at our maximum field $B = 12$ T. The hole resistance $R_h$ remains above our measurement range of 1 MΩ, and the drag ratio remains close to unity with no noticeable QOs. At higher $V_B$, the exciton binding energy is reduced by many-body interaction effects, and QOs become obvious when the magnetic energy scale competes with it at high field. For example, for $V_B = 0.41$ V well-defined QOs can be observed in both drag ratio and $R_h$, which are periodic with the inverse field $1/B$. The oscillation magnitude grows larger at higher $B$ due to increased cyclotron energy. For $V_B = 0.43$ V, the drag ratio oscillates to zero at integer LL filling, signaling a quantum phase transition from an EI into two independent QHIs.

## Finite-temperature phase diagram

Finally, we explore the phase diagram of the e-h fluids as a function of the pair density and temperature. Fig. 4a displays the temperature dependence of the drag ratio at $B = 12$ T along the CNL. Multiple EI domes appear at low temperatures. They are roughly centered around half-integer LL fillings up to a filling factor of 6. With increased temperature, the drag ratio sharply drops to zero for high-order EI domes, while a smoother transition is observed for lower order ones (Fig. 4b). The exciton melting temperature being lower for large filling factors is consistent with a reduction of exciton binding energy when both electrons and holes are limited to high LLs by Landau quantization. The corresponding hole resistance at half-integer fillings (Fig. 4c-d) decreases with increasing temperature, corroborating the insulating nature of these EI phases.

Interestingly, we observe temperature-induced EI-QHI phase transitions at small integer filling factors. As seen in Fig. 4a and Fig. 2e, the drag ratio dips at $\nu = 1 - 4$ does not decrease to zero at our lowest temperature due to finite LL mixing induced by the Coulomb interactions. Fig. 4e-f shows the temperature dependence of $R_h$ and the drag ratio at integer LL fillings $\nu = 3$ and 4. The drag ratio shows a rapid decrease as the temperature is increased beyond ~5 K for $\nu = 3$ and ~2 K for $\nu = 4$. Meanwhile, the deviation of $R_h$ from $h/\nu e^2$ also disappears, recovering the quantized values in the QHI. This suggests temperature-induced phase transitions, where thermal energy melts excitons in the EI and recovers the QHI phase.

In summary, we have reported QOs and EI-QHI phase transitions in strongly interacting e-h bilayers. The transport behavior of e-h fluids under a strong magnetic field reveals rich and complex phases, offering insights into the interplay between the magnetic field and e-h interactions. The unambiguous observation of QOs with $1/B$ periodicity in both drag signal and $R_h$ connects with previous theoretical predictions, confirming that QOs can arise in EIs. Our results also establish TMD-based e-h bilayers as a promising platform for studying exciton physics in the QH regime.

## Methods

**Device fabrication.** We use a dry-transfer method based on polyethylene terephthalate glycol (PETG) stamps to fabricate the heterostructures. Monolayer $MoSe_2$, monolayer $WSe_2$, few-layer graphene and hBN flakes are mechanically exfoliated from bulk crystals onto $SiO_2$/Si substrates. We use 6-10 nm hBN as the gate dielectric and 2-3 nm thin hBN as the interlayer spacer. Prior to the stacking process, metal electrodes for the hole layer (7 nm Pt with 3 nm Cr adhesion layer) are defined using photolithography (Durham Magneto Optics, MicroWriter) and electron beam evaporation (Angstrom Engineering) onto a high resistivity SiO$_2$/Si substrate. The graphite top gate is cut using an atomic force microscope tip[43] to eliminate undesired current paths in the unmatched WSe$_2$ monolayer regions.

A 0.5 mm thick clear PETG stamp is employed to pick up the flakes at 65-75 °C. A >100 nm thick hBN is first picked up by the stamp to protect the following layers. Then the graphite top gate, the top dielectric hBN, the monolayer WSe$_2$, the thin hBN spacer, the two thicker hBN spacers in the reservoir regions, the two graphite contacts, the monolayer MoSe$_2$, the bottom dielectric hBN, and the bottom graphite gate are sequentially picked up. The whole stack is released onto the prepatterned Pt electrodes at 100 °C, followed by dissolving the PETG in chloroform for one day. Finally, metal electrodes (5 nm Cr/60 nm Au) are defined using photolithography and electron beam evaporation.

**Optical measurements.** The optical measurements are performed in an optical cryostat (Quantum Design, OptiCool) with a nominal base temperature of 2 K. The reflection spectroscopy is performed with a supercontinuum laser (Fianium Femtopower 1060) as the light source. The laser is focused on the sample by a 20× Mitutoyo objective with ~1.5 μm beam diameter. The reflected light is collected after a spectrometer by a CCD camera (Princeton Instruments PIXIS 256e) with 1000 ms exposure time. We use reflection spectroscopy to determine the electron and hole densities shown in Fig. 1c-d. A detailed description of the methodology can be found in ref.[1].

**WSe$_2$ hole resistance measurements.** The transport measurements are performed in a dilution refrigerator (Bluefors LD250) with a base lattice temperature of 10 mK. All the signal wires are filtered at the mixing chamber flange by $RC$ and $RL$ filters (QDevil) before reaching the sample. The electronic temperature, calibrated by the $1/T$ dependence of spin susceptibility of a 2DEG in the Wigner crystal regime (measured by the magnetic circular dichroism signal in a monolayer MoSe$_2$ device)[44], is about 0.5 K. The d.c. voltages on the device are applied with Keithley 2400/2450 source meters or Keithley 2502 picoammeters.

The detailed circuit diagram for the two-terminal WSe$_2$ hole resistance measurements is shown in Extended Data Fig. 5a. Good electrical contacts are crucial for accurate two-terminal resistance measurements of quantum Hall devices. In our device geometry, the contact resistance includes two parts: the resistance at Pt to WSe$_2$ junction, and resistance at the reservoir region to region of interest junction. Since we have multiple Pt electrodes in each reservoir region, we exclude the Pt-WSe$_2$ junction resistance using two separate pairs of Pt electrode for sourcing the current and measuring the voltage. A current excitation at 17 Hz is applied between two WSe$_2$ electrodes, and the voltage drop between another two electrodes is used to determine the hole layer resistance. However, note that multiple contacts are connected to the same highly doped reservoir region. It does not exclude the contact resistance from the reservoirs to the region of interest. Therefore, the measured $R_h$ should be interpreted as two-terminal resistance in the QH regime[40], where each highly doped reservoir region acts as one terminal.

The Pt-WSe$_2$ contact resistance can be determined by comparing the resistance using two different measurement configurations shown in Extended Data Fig. 3a. In configuration I, all three Pt electrodes in each reservoir region are linked together to source currents and measure the voltage drop. The resulting two-terminal resistance includes the Pt-WSe$_2$ contact resistance. In configuration II (the configuration used in the main text), different pairs of Pt electrodes are used for current excitation and voltage measurement, excluding the Pt-WSe$_2$ contact resistance.

Extended Data Fig. 3b shows the measured resistance of the two configurations. Their difference is around 2 kΩ and depends weakly on the magnetic field.

**Coulomb drag measurements.** Extended Data Fig. 5b shows the detailed circuit diagram of the Coulomb drag measurements. Closed-circuit drag measurements require good electrical contacts for the drag layer. The graphite-$MoSe_2$ contact resistance is 2-3 MΩ, much larger than the Pt-$WSe_2$ contact. We therefore choose the $MoSe_2$ layer as the drive layer and the $WSe_2$ layer as the drag layer. A 5 $mV_{rms}$ a.c. voltage excitation at 17 Hz is applied between the two $MoSe_2$ contacts. A 10 kΩ potentiometer is used to distribute the a.c. voltage between the two contacts to minimize interlayer capacitive coupling. The drag current is measured between the left three $WSe_2$ electrodes and the right three $WSe_2$ electrodes.

The interlayer tunnelling current is at picoampere level (Extended Data Fig. 6), providing a lower bound of the tunnelling resistance $> 10^{10}$ Ω. This is many orders of magnitude larger than the in-plane resistance, and therefore tunneling does not contribute to the measured drive and drag currents.

# References

1. Qi, R. *et al.* Thermodynamic behavior of correlated electron-hole fluids in van der Waals heterostructures. *Nat. Commun.* **14**, 8264 (2023).

2. Ma, L. *et al.* Strongly correlated excitonic insulator in atomic double layers. *Nature* **598**, 585–589 (2021).

3. Nguyen, P. X. *et al.* Perfect Coulomb drag in a dipolar excitonic insulator. Preprint at https://doi.org/10.48550/arXiv.2309.14940 (2023).

4. Qi, R. *et al.* Perfect Coulomb drag and exciton transport in an excitonic insulator. Preprint at https://doi.org/10.48550/arXiv.2309.15357 (2023).

5. Shao, Y. & Dai, X. Quantum oscillations in an excitonic insulating electron-hole bilayer. *Phys. Rev. B* **109**, 155107 (2024).

6. Zou, B., Zeng, Y., MacDonald, A. H. & Strashko, A. Electrical control of two-dimensional electron-hole fluids in the quantum Hall regime. *Phys. Rev. B* **109**, 085416 (2024).

7. Zeng, Y. & MacDonald, A. H. Electrically controlled two-dimensional electron-hole fluids. *Phys. Rev. B* **102**, 085154 (2020).

8. Wu, F.-C., Xue, F. & MacDonald, A. H. Theory of two-dimensional spatially indirect equilibrium exciton condensates. *Phys. Rev. B* **92**, 165121 (2015).

9. De Palo, S., Rapisarda, F. & Senatore, G. Excitonic Condensation in a Symmetric Electron-Hole Bilayer. *Phys. Rev. Lett.* **88**, 206401 (2002).

10. Fogler, M. M., Butov, L. V. & Novoselov, K. S. High-temperature superfluidity with indirect excitons in van der Waals heterostructures. *Nat. Commun.* **5**, 4555 (2014).

11. Liu, X., Watanabe, K., Taniguchi, T., Halperin, B. I. & Kim, P. Quantum Hall drag of exciton condensate in graphene. *Nat. Phys.* **13**, 746–750 (2017).

12. Eisenstein, J. P. & MacDonald, A. H. Bose–Einstein condensation of excitons in bilayer electron systems. *Nature* **432**, 691–694 (2004).

13. Zhu, X., Littlewood, P. B., Hybertsen, M. S. & Rice, T. M. Exciton Condensate in Semiconductor Quantum Well Structures. *Phys. Rev. Lett.* **74**, 1633–1636 (1995).

14. Maezono, R., López Ríos, P., Ogawa, T. & Needs, R. J. Excitons and biexcitons in symmetric electron-hole bilayers. *Phys. Rev. Lett.* **110**, 216407 (2013).

15. Dai, D. D. & Fu, L. Strong-Coupling Phases of Trions and Excitons in Electron-Hole Bilayers at Commensurate Densities. *Phys. Rev. Lett.* **132**, 196202 (2024).

16. Qi, R. *et al.* Electrically controlled interlayer trion fluid in electron-hole bilayers. Preprint at https://doi.org/10.48550/arXiv.2312.03251 (2023).

17. Nguyen, P. X. *et al.* A degenerate trion liquid in atomic double layers. Preprint at https://doi.org/10.48550/arXiv.2312.12571 (2023).

18. Liu, X. *et al.* Crossover between strongly coupled and weakly coupled exciton superfluids. *Science* **375**, 205–209 (2022).

19. Eisenstein, J. P. Exciton Condensation in Bilayer Quantum Hall Systems. *Annu. Rev. Condens. Matter Phys.* **5**, 159–181 (2014).

20. Li, J. I. A., Taniguchi, T., Watanabe, K., Hone, J. & Dean, C. R. Excitonic superfluid phase in double bilayer graphene. *Nat. Phys.* **13**, 751–755 (2017).

21. Croxall, A. F. *et al.* Anomalous Coulomb Drag in Electron-Hole Bilayers. *Phys. Rev. Lett.* **101**, 246801 (2008).

22. Du, L. *et al.* Evidence for a topological excitonic insulator in InAs/GaSb bilayers. *Nat. Commun.* **8**, 1971 (2017).

23. Wang, R., Sedrakyan, T. A., Wang, B., Du, L. & Du, R.-R. Excitonic topological order in imbalanced electron–hole bilayers. *Nature* **619**, 57–62 (2023).

24. Han, Z., Li, T., Zhang, L., Sullivan, G. & Du, R.-R. Anomalous Conductance Oscillations in the Hybridization Gap of $\mathrm{InAs}/\mathrm{GaSb}$ Quantum Wells. *Phys. Rev. Lett.* **123**, 126803 (2019).

25. Xiao, D., Liu, C.-X., Samarth, N. & Hu, L.-H. Anomalous Quantum Oscillations of Interacting Electron-Hole Gases in Inverted Type-II $\mathrm{InAs}/\mathrm{GaSb}$ Quantum Wells. *Phys. Rev. Lett.* **122**, 186802 (2019).

26. Li, L., Sun, K., Kurdak, C. & Allen, J. W. Emergent mystery in the Kondo insulator samarium hexaboride. *Nat. Rev. Phys.* **2**, 463–479 (2020).

27. Pirie, H. *et al.* Visualizing the atomic-scale origin of metallic behavior in Kondo insulators. *Science* **379**, 1214–1218 (2023).

28. Shen, H. & Fu, L. Quantum Oscillation from In-Gap States and a Non-Hermitian Landau Level Problem. *Phys. Rev. Lett.* **121**, 026403 (2018).

29. Zhang, L., Song, X.-Y. & Wang, F. Quantum Oscillation in Narrow-Gap Topological Insulators. *Phys. Rev. Lett.* **116**, 046404 (2016).

30. Knolle, J. & Cooper, N. R. Quantum Oscillations without a Fermi Surface and the Anomalous de Haas-van Alphen Effect. *Phys. Rev. Lett.* **115**, 146401 (2015).

31. Knolle, J. & Cooper, N. R. Excitons in topological Kondo insulators: Theory of thermodynamic and transport anomalies in SmB6. *Phys. Rev. Lett.* **118**, 096604 (2017).

32. Erten, O., Chang, P.-Y., Coleman, P. & Tsvelik, A. M. Skyrme Insulators: Insulators at the Brink of Superconductivity. *Phys. Rev. Lett.* **119**, 057603 (2017).

33. Lee, P. A. Quantum oscillations in the activated conductivity in excitonic insulators: Possible application to monolayer WTe2. *Phys. Rev. B* **103**, L041101 (2021).

34. He, W.-Y. & Lee, P. A. Quantum oscillation of thermally activated conductivity in a monolayer WTe2-like excitonic insulator. *Phys. Rev. B* **104**, L041110 (2021).

35. Chowdhury, D., Sodemann, I. & Senthil, T. Mixed-valence insulators with neutral Fermi surfaces. *Nat. Commun.* **9**, 1766 (2018).

36. Sodemann, I., Chowdhury, D. & Senthil, T. Quantum oscillations in insulators with neutral Fermi surfaces. *Phys. Rev. B* **97**, 045152 (2018).

37. Li, G. *et al.* Two-dimensional Fermi surfaces in Kondo insulator SmB6. *Science* **346**, 1208–1212 (2014).

38. Tan, B. S. *et al.* Unconventional Fermi surface in an insulating state. *Science* **349**, 287–290 (2015).

39. Wang, P. *et al.* Landau quantization and highly mobile fermions in an insulator. *Nature* **589**, 225–229 (2021).

40. Rikken, G. L. J. A. *et al.* Two-terminal resistance of quantum Hall devices. *Phys. Rev. B* **37**, 6181–6186 (1988).

41. Zhu, J., Li, T., Young, A. F., Shan, J. & Mak, K. F. Quantum Oscillations in Two-Dimensional Insulators Induced by Graphite Gates. *Phys. Rev. Lett.* **127**, 247702 (2021).

42. Fenton, E. W. Excitonic Insulator in a Magnetic Field. *Phys. Rev.* **170**, 816–821 (1968).

43. Li, H. *et al.* Electrode-Free Anodic Oxidation Nanolithography of Low-Dimensional Materials. *Nano Lett.* **18**, 8011–8015 (2018).

44. Sung, J. *et al.* Observation of an electronic microemulsion phase emerging from a quantum crystal-to-liquid transition. Preprint at https://doi.org/10.48550/arXiv.2311.18069 (2023).

**Data availability**
The data that support the findings of this study are available from the corresponding author upon request.

**Acknowledgements**
This work was primarily funded by the U.S. Department of Energy, Office of Science, Basic Energy Sciences, Materials Sciences and Engineering Division under Contract No. DE-AC02-05-CH11231 within the van der Waals heterostructure program KCFW16 (device fabrication, transport measurements). Optical characterization of the excitonic insulator was supported by the AFOSR award FA9550-23-1-0246. K.W. and T.T. acknowledge support from the Japan Society for the Promotion of Science (KAKENHI grants 21H05233 and 23H02052) and World Premier International Research Center Initiative, Ministry of Education, Culture, Sports, Science and Technology, Japan. R.Q. acknowledges support from Kavli ENSI graduate student fellowship.

**Author contributions**
F.W. and R.Q. conceived the research. Q.L and R.Q. fabricated the devices with help from Z.C and S.C. R.Q. performed the optical measurements assisted by Z.Z. and J.X. R.Q. and Z.Z. performed the transport measurements with help from H.K. and C.S. R.Q., Z.Z. and F.W. analyzed the data with inputs from M.F.C., B.Z. and A.H.M. K.W. and T.T. grew hBN crystals. All authors discussed the results and wrote the manuscript.

**Competing interests**
The authors declare no competing interests.

# Figures

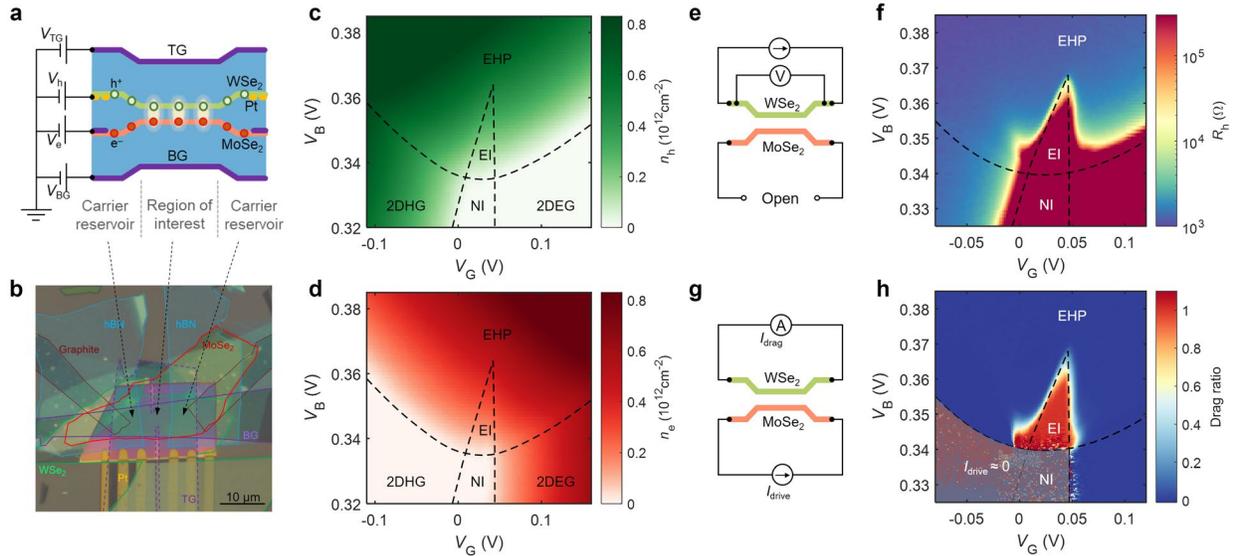

**Fig. 1 | Strongly coupled electron-hole bilayers.**

**a**, Schematic cross-section of the electron-hole bilayer device based on the MoSe$_2$/hBN/WSe$_2$ heterostructure.

**b**, An optical microscopy image of device D1 (WSe$_2$/7-layer hBN/MoSe$_2$ heterostructure), with flake boundaries outlined.

**c-d**, Experimentally determined hole density $n_h$ in the WSe$_2$ layer (**c**) and electron density $n_e$ in the MoSe$_2$ layer (**d**), respectively, measured at zero magnetic field and a nominal temperature of 2 K. The phase diagram is composed of normal insulator (NI), excitonic insulator (EI), 2D electron gas (2DEG), 2D hole gas (2DHG), and electron-hole plasma (EHP) regions. The vertical electric field is fixed by antisymmetric gating $V_{BG} - V_{TG} = 7$ V while the gate voltage $V_G$ and interlayer bias voltage $V_B$ are varied.

**e**, Circuit diagram of the two-terminal hole resistance measurement. The resistance in the WSe$_2$ layer is measured with the MoSe$_2$ layer open circuit.

**f**, 2D color plot of the hole resistance $R_h$ as a function of $V_G$ and $V_B$ at $T = 0.5$ K. The EI region exhibits a large $R_h$ although the hole density is finite.

**g**, Circuit diagram of the closed-circuit Coulomb drag measurement. A drive current is passed through the MoSe$_2$ layer, and the closed-circuit drag current in the WSe$_2$ layer is measured.

**h**, 2D color plot of the drag ratio $I_{drag}/I_{drive}$ as a function of $V_G$ and $V_B$ at $T = 0.5$ K. Perfect Coulomb drag is observed in the EI region. The bottom-left region shaded with semitransparent gray color is noisy due to the absence of any drive current when $n_e = 0$.

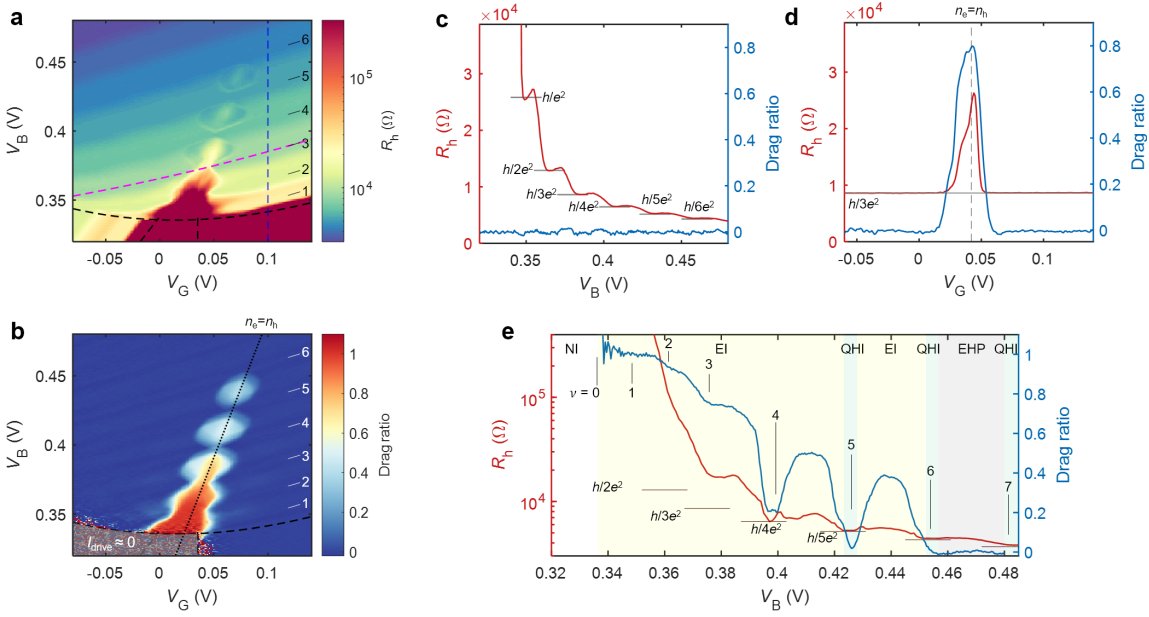

**Fig. 2 | Electron-hole fluids in a strong magnetic field.**

**a**, Measured hole resistance $R_h$ as a function of $V_G$ and $V_B$ at $B = 12$ T. Landau quantization of holes leads to the quasi-periodic structures following $n_h$ contour lines outside the EI region. The EI region develops into a series of lobes centered near the CNL and exhibits increased $R_h$ values.

**b**, Closed-circuit drag ratio $I_{\text{drag}}/I_{\text{drive}}$ at $B = 12$ T. Strong Coulomb drag is observed in the EI region near the CNL, which matches the EI region in **a**. Other regions show negligible drag signal.

**c**, $V_B$ dependence of $R_h$ (left axis) and the drag ratio (right axis) at fixed gating $V_G = 0.10$ V (blue dashed line in **a**), where the electron and hole densities are not equal. At integer hole filling factors, $R_h$ quantizes at $h/Ne^2$, indicating quantum Hall states of holes.

**d**, Line cut of $R_h$ (left axis) and the drag ratio (right axis) along a fixed hole LL filling $\nu_h = 3$ (magenta dashed line in **a**). $R_h$ deviates from $h/3e^2$ only near the net charge neutrality, accompanied by a simultaneous increase in the drag signal.

**e**, Line cut of $R_h$ (left axis) and the drag ratio (right axis) along the net CNL $n_e = n_h$ (black dotted line in **b**). Clear QOs can be observed in both $R_h$ and drag signals. Multiple EI-QHI phase transitions happen at higher $V_B$, where the exciton binding energy and the cyclotron energy are comparable and compete with each other. The EI regions are characterized by a large drag signal and enhanced $R_h$, while the QHI regions exhibit quantized $R_h$ values and very small drag signal. The EI, QHI, and EHP regions are highlighted with yellow, green, and grey shade, respectively.

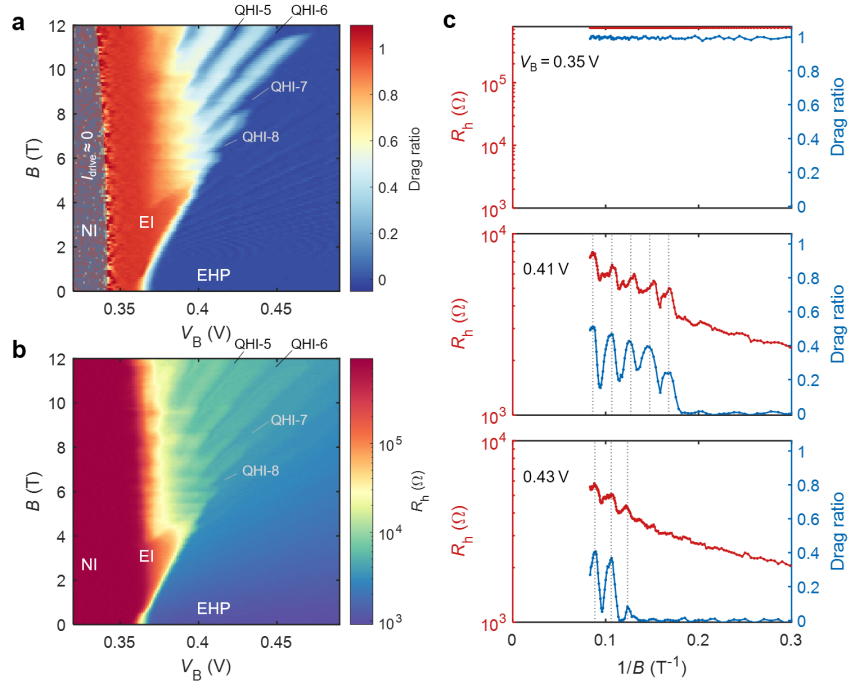

**Fig. 3 | Quantum oscillations at net charge neutrality.**

**a**, Closed-circuit drag ratio as a function of perpendicular magnetic field $B$ and interlayer bias voltage $V_B$. QOs in the EI phase start to develop at $B \gtrsim 5$ T, and evolve into EI-QHI phase transitions at higher $B$.

**b**, $R_h$ as a function of $B$ and $V_B$. The Landau-fan-like structures match with corresponding features in **a**.

**c**, Magnetic field dependence of $R_h$ (left axis) and the drag ratio (right axis) at three typical $V_B$ values. At small $V_B$ (top panel), no QO is observed due to the large exciton binding energy at low density. At higher $V_B$ (middle and bottom panels), clear QOs with $1/B$ periodicity show up in both $R_h$ and drag signals.

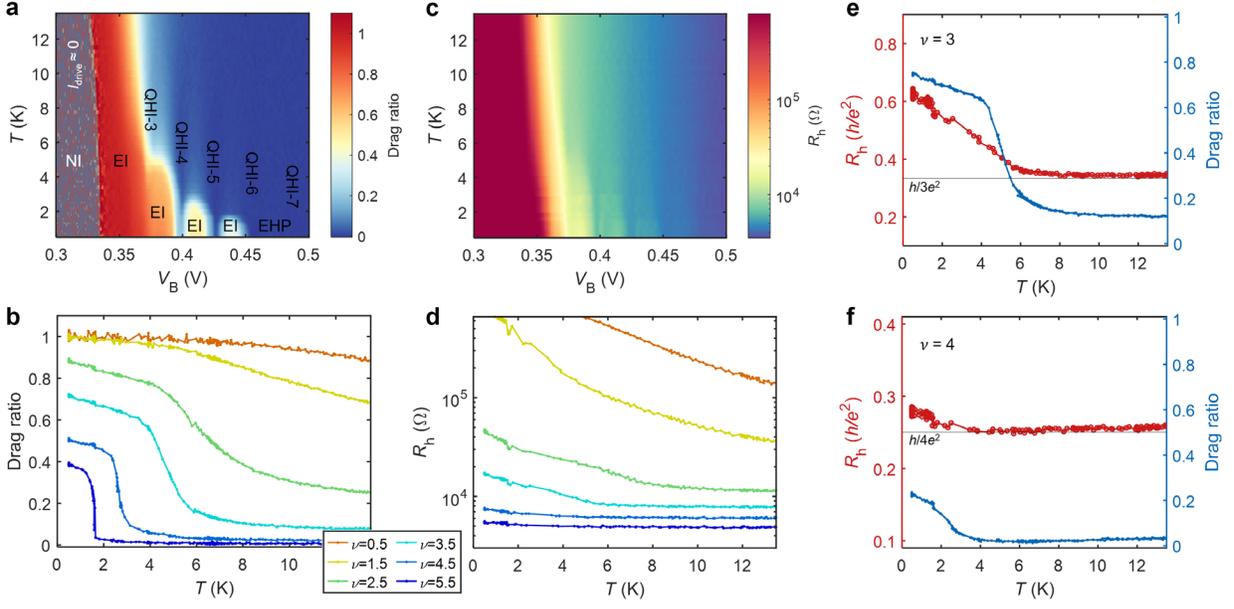

**Fig. 4 | Finite temperature phase diagram at net charge neutrality.**

**a**, Drag ratio along the CNL at $B = 12$ T for different $V_B$ and temperatures. There are multiple EI domes centered around half-integer fillings. High-order EI regions are separated by QHI states, while the lower ones are connected. The exciton ionization temperature decreases with the exciton density (i.e. higher $V_B$).

**b**, Linecuts of the drag ratio at different half-integer LL fillings as a function of temperature.

**c**, $R_h$ along the CNL at $B = 12$ T for different $V_B$ and temperatures. $R_h$ decreases with increasing temperature in the EI phases, confirming the insulating behavior.

**d**, Linecuts of $R_h$ at half-integer LL fillings as a function of temperature.

**e-f**, $R_h$ (left axis) and the drag ratio (right axis) at integer LL filling $\nu_e = \nu_h = 3$ (**e**) and 4 (**f**) as a function of temperature. The strong decrease of drag signals and the recovery of $R_h$ quantization at elevated temperature suggests thermally driven EI-QHI phase transitions, in which thermal energy assists the cyclotron energy to dissociate exciton pairing.

# Extended Data Figures

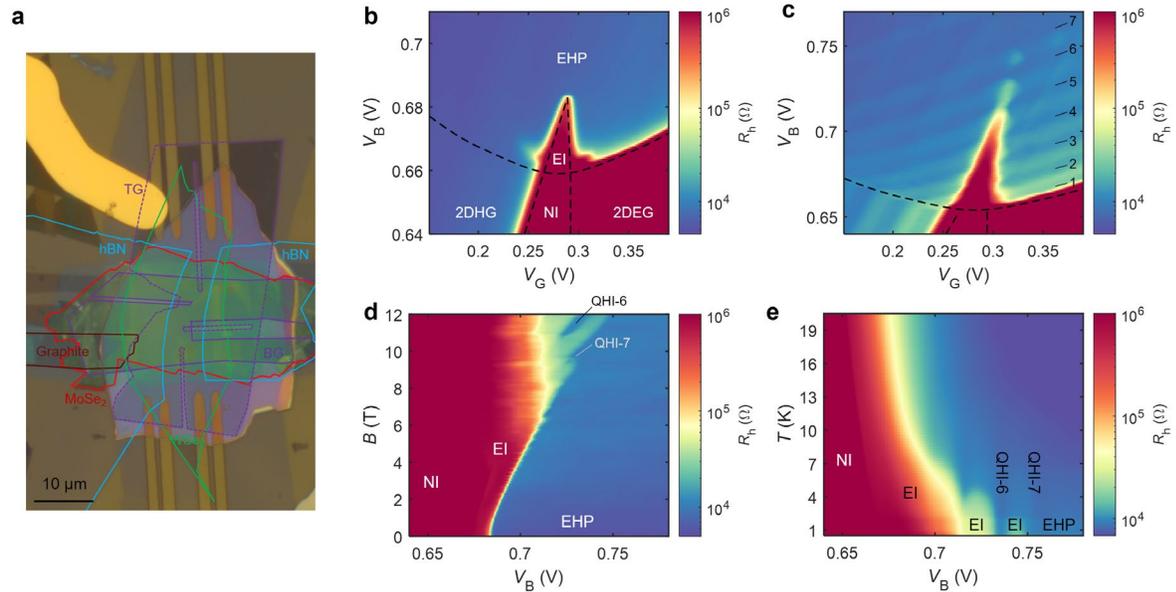

**Extended Data Fig. 1 | Data from a second device.**

**a**, An optical microscopy image of device D2 (WSe$_2$/6-layer hBN/MoSe$_2$ heterostructure) with flakes outlined. In this device, the MoSe$_2$ contacts in the second reservoir region are mechanically broken, preventing us from performing the Coulomb drag measurements.

**b-c**, Measured $R_h$ at $B = 0$ T (**b**) and 12 T (**c**). The results are qualitatively consistent with the data from device D1 shown in the main text.

**d**, Fan diagram of $R_h$ along the CNL. Due to a smaller interlayer distance compared to device D1, the exciton binding energy is larger, requiring a larger magnetic field to induce QOs and EI-QHI phase transitions.

**e**, $R_h$ along the CNL at $B = 12$ T as a function of $V_B$ and $T$.

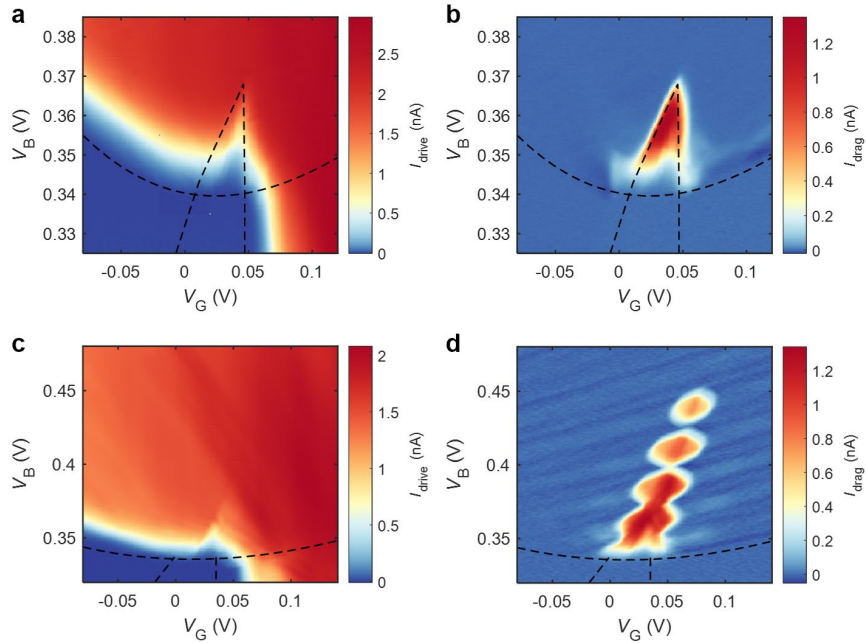

**Extended Data Fig. 2 | Drive and drag currents.**

**a-b**, Drive and drag currents at $B = 0$ with 5 mV drive voltage applied across the MoSe$_2$ layer.

**c-d**, Drive and drag currents at $B = 12$ T.

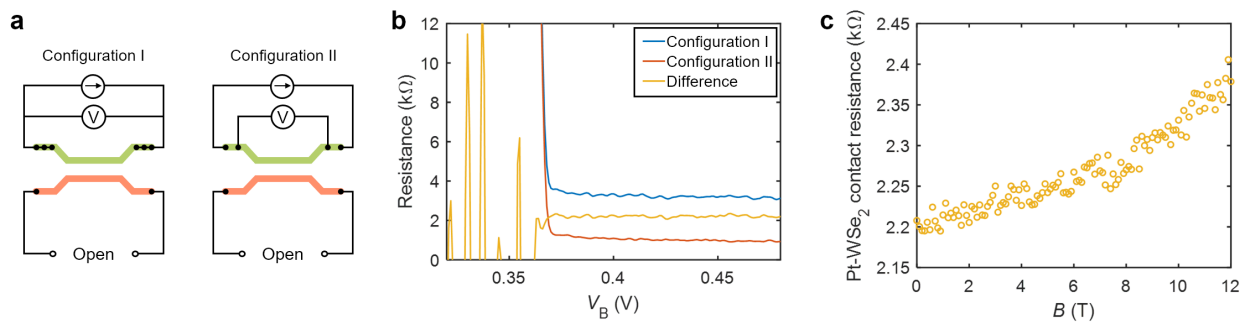

**Extended Data Fig. 3 | Determination of Pt-WSe$_2$ contact resistance.**

**a,** Schematic circuit diagram for two different $R_h$ measurement configurations.

**b,** Comparison of the measured $R_h$ as a function of $V_B$ along the CNL, measured at $T = 0.5$ K, $B = 0$. Their difference (yellow line) corresponds to the total Pt-WSe$_2$ contact resistance on the two sides. Nearly no $V_B$ dependence is observed, with an average value of 2.2 kΩ.

**c,** Magnetic field dependence of the Pt-WSe$_2$ contact resistance, showing a slight increase at high fields.

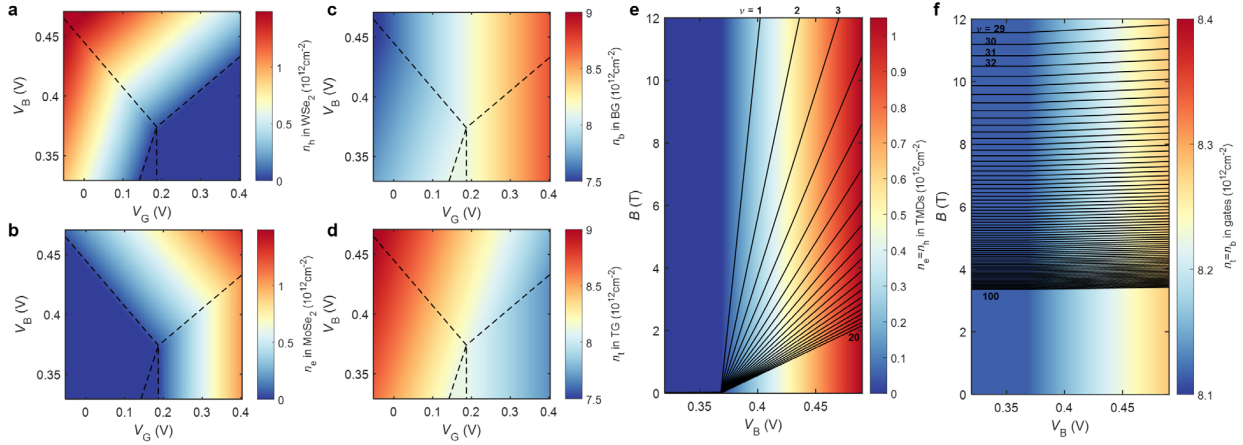

**Extended Data Fig. 4 | Carrier densities in graphite gates modeled by a capacitor model.**

**a-d**, The hole density $n_h$ in the WSe$_2$ layer (**a**), the electron density $n_e$ in the MoSe$_2$ layer (**b**), the electron density in the graphite top gate $n_t$ (**c**), and the hole density in the graphite bottom gate $n_b$ (**d**), as a function of $V_G$ and $V_B$. The densities are calculated based on a simple capacitor model that accounts for the geometric capacitances and the type-II band gap (1.5 eV, assumed to be symmetric with respect to the graphite Fermi level)[2]. Parameters used in the model: hBN out-of-plane dielectric constant 3.5, top and bottom hBN thicknesses 7 nm, interlayer distance 2.7 nm, antisymmetric gating $V_{BG} - V_{TG} = 7$ V.

**e-f**, Calculated densities in the TMD layers (**e**) and the graphite gates (**f**) as a function of $V_B$ along the net charge neutrality. Solid lines trace integer LL fillings in the magnetic field. The densities in the gates are always about 10 times higher than that in the TMD layers, and depend on $V_B$ very weakly. Therefore, the QOs in the gates would happen at a dramatically higher frequency along the $B$ axis and cannot explain the experimental data.

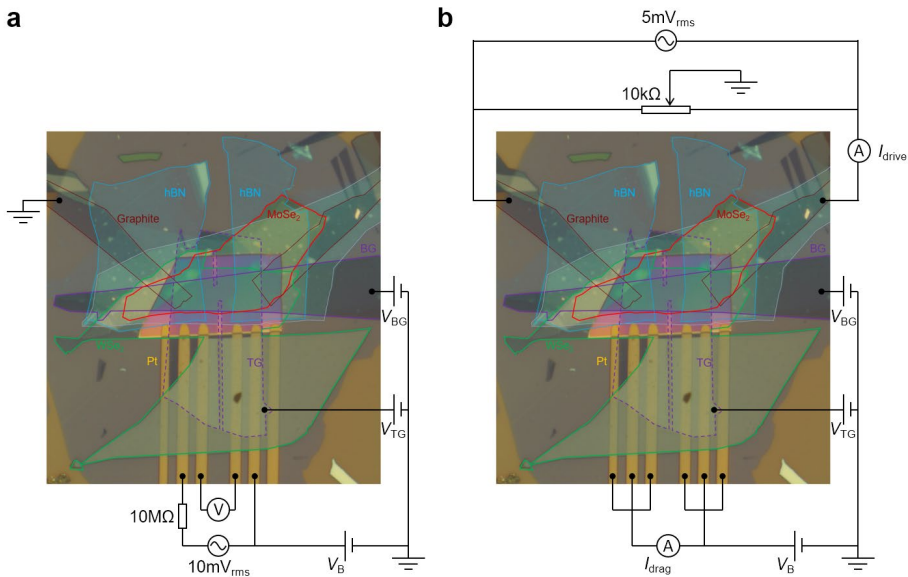

**Extended Data Fig. 5 | Circuit diagram for the transport measurements.**

**a**, Circuit diagram for two-terminal $R_h$ measurement.

**b**, Circuit diagram for the closed-circuit Coulomb drag measurement.

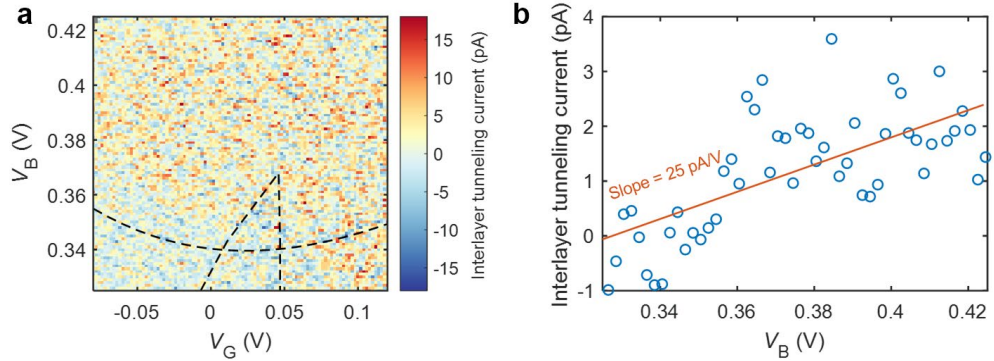

**Extended Data Fig. 6 | Interlayer tunneling current.**

**a**, Interlayer tunneling current as a function of $V_G$ and $V_B$ at $T = 0.5$ K, $B = 0$. The tunneling current is barely above the picoampere noise level.

**b**, A linecut of tunneling current along net charge neutrality $n_e = n_h$ as a function of $V_B$. The experimental data (open circles) is fitted to a linear line to estimate the interlayer tunneling resistance, giving a lower bound of 40 GΩ.